\documentclass{article}
\usepackage{spconf}
\usepackage{xcolor}
\usepackage{amsmath}
\usepackage{amsfonts}
\usepackage{caption}
\usepackage{colortbl}
\usepackage{booktabs}
\usepackage{multirow}
\usepackage{subcaption}
\usepackage{graphicx}
\usepackage{setspace}
\usepackage{cite}
\usepackage{tikz}
\usetikzlibrary{backgrounds}

\newcommand{\mypar}[1]{{\bf #1.}}

\definecolor{darkblue}{RGB}{0,0,100}	
\definecolor{darkredc}{RGB}{90,0,0}
\definecolor{thisblue}{rgb}{0.03, 0.27, 0.49}
\definecolor{darkgreen}{rgb}{0.0, 0.42, 0.24}
\definecolor{red}{RGB}{163,0,0}

\allowdisplaybreaks

\title{Enhanced Hyperspectral Image Super-Resolution via RGB Fusion and TV-TV Minimization}
\name{Author(s) Name(s)\thanks{Thanks to XYZ agency for funding.}}
\address{Author Affiliation(s)}
\name{Marija Vella$^{\star }$,\, Bowen Zhang$^{\star \ast}$, \, Wei Chen$^{\star \ast}$, \, Jo\~{a}o F. C. Mota$^{\star}$ }
\address{ $^{\star}$ Institute of Sensors, Signals and Systems, Heriot-Watt
University, Edinburgh EH14 4AS, UK\\ $^{\star \ast}$ State Key Laboratory of Rail Traffic Control and Safety, Beijing Jiaotong University, China}
\begin{document}
\ninept
\maketitle
\begin{abstract}
Hyperspectral (HS) images contain detailed spectral information that has proven crucial in applications like remote sensing, surveillance, and astronomy. However, because of hardware limitations of HS cameras, the captured images have low spatial resolution. To improve them, the low-resolution hyperspectral images are fused with conventional high-resolution RGB images via a technique known as fusion based HS image super-resolution. Currently, the best performance in this task is achieved by deep learning (DL) methods. Such methods, however, cannot guarantee that the input measurements are satisfied in the recovered image, since the learned parameters by the network are applied to every test image. Conversely, model-based algorithms can typically guarantee such measurement consistency. Inspired by these observations, we propose a framework that integrates learning and model based methods. Experimental results show that our method produces images of superior spatial and spectral resolution compared to the current leading methods, whether model- or DL-based.
\end{abstract}
\section{Introduction}
\label{sec:intro}

Hyperspectral (HS) cameras sense the electromagnetic spectrum to produce images
that depict a scene across several contiguous bands. Such technology has led to
improved performance in various tasks, including anomaly detection~\cite{Xu2016Anomaly} and remote sensing~\cite{Ojha2015spectral}. 

HS cameras, however, suffer from low spatial and temporal resolution
\cite{Schechner2002generalized}. RGB cameras, on the other hand, acquire images
with high spatial resolution but low spectral resolution, since they split the
spectrum into three broad bands. This tradeoff between spatial and spectral
resolution stems from the physics of the acquisition process. To overcome it,
one can fuse a low (spatial) resolution hyperspectral (LRHS) image with a high resolution (spatial) RGB/multispectral (HRMS) image to obtain the high resolution hyperspectral (HRHS) image, a method known as fusion based HS image super-resolution.

\mypar{Acquisition model}
Let $\boldsymbol{X} \in \mathbb{R}^{M_0 \cdot N_0 \times S_0}$ denote the unknown HRHS that we wish to reconstruct. Each column represents a vectorized HRHS of size $M_0 \times N_0$ on a given spectral band, and there are $S_0$ spectral bands. We assume access to a low-resolution HS image acquired by a HS camera, $\boldsymbol{Z} \in \mathbb{R}^{M\cdot N \times S_0}$, where $M < M_0$ and $N < N_0$, and to a high-resolution RGB image acquired by a conventional camera, $\boldsymbol{Y} \in \mathbb{R}^{M_0 \cdot N_0 \times S}$, where $S<S_0$. Typically, $S = 3$ and $S_0$ is greater than 20. All these quantities are related as

\begin{equation}
\boldsymbol{Z} = \boldsymbol{AX}, \quad \boldsymbol{Y}=\boldsymbol{XR},
\label{eq:cond1}
\end{equation}
where $\boldsymbol{A} \in \mathbb{R}^{M \cdot N \times M_0 \cdot N_0}$ is the downscaling operator, and $\boldsymbol{R} \in \mathbb{R}^{S_0 \times S}$ is the camera spectral response (CSR) function of the RGB camera i.e., it integrates all bands of the spectrum into the RGB image.  

\begin{figure}[t]
	\includegraphics[width=8.5cm, height=4.3cm]{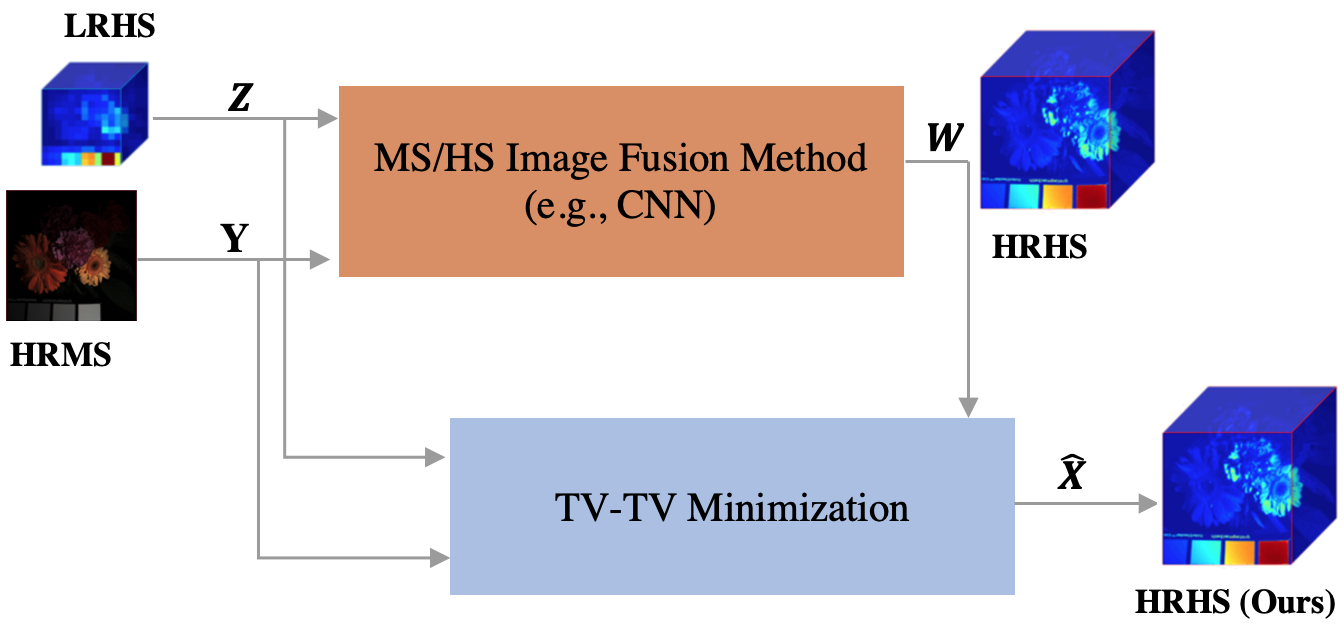}
	\caption{Our framework. A high-resolution RGB image $\boldsymbol{Y}$ (HRMS) and a low-resolution HS image $\boldsymbol{Z}$ (LRHS) are first super-resolved into $\boldsymbol{W}$ with a fusion-based hyperspectral imaging method (typically, a convolutional neural network). We then create a measurement-consistent high-resolution HS image $\boldsymbol{\widehat{X}}$ (HRHS) by solving TV-TV minimization using $\boldsymbol{Z}$, $\boldsymbol{Y}$, and $\boldsymbol{W}$ as inputs.}
	\label{fig:frameworkV2}
	\vspace{-0.4cm}
\end{figure}

\mypar{Overview of existing approaches} Reconstructing $\boldsymbol{X}$ from $\boldsymbol{Z}$ and $\boldsymbol{Y}$ in (\ref{eq:cond1}) is an ill-posed problem, as there are generally more unknowns than equations. This task thus requires assumptions about the structure of $\boldsymbol{X}$ (prior knowledge). Examples include 
sparsity \cite{Akhtar2014sparsespatiospectral} and spectral correlation \cite{nezhad2016fusion}. Other approaches to recover $\boldsymbol{X}$ encode prior knowledge about the HS observations via matrix \cite{Kawakami2011tfm, Yokoya2012cnmf} and tensor \cite{dian2017nlstf} factorization. We refer to this class of methods, in which prior knowledge is explicitly encoded, as \textit{model-based} methods. 

Deep learning (DL) methods have also been applied to reconstruct $\boldsymbol{X}$ from $\boldsymbol{Z}$ and $\boldsymbol{Y}$, e.g., \cite{xie2019fusion, qu2018unsupervised}. Although they require extensive training datasets, they bypass the need to encode assumptions explicitly and are computationally efficient once trained. Currently, they are the state-of-the-art for this task. DL methods, however, suffer from two important drawbacks: they generally lack interpretability, and during testing, they are unable to enforce measurement consistency between the inputs ($\boldsymbol{Z}, \boldsymbol{Y}$) and their output, i.e., enforce (\ref{eq:cond1}) after they have been trained. Conventional model-based algorithms do not suffer from these drawbacks, but produce outputs of lower quality.

\mypar{Problem statement} Our aim is then to design a method for fusion based HS image super-resolution that not only produces high quality outputs, but also guarantees measurement consistency for every test image. By guaranteeing such consistency, we aim to preserve fine details in the measurements, which can be critical in applications like airport security and space surveillance.

We propose to combine DL (or any other RGB-guided HS image super-resolution method) with model-based methods. Fig. \ref{fig:frameworkV2} illustrates our scheme, described in Section \ref{ourframework}. In brief, the inputs $\boldsymbol{Y}$ and $\boldsymbol{Z}$ are passed through a fusion based super-resolution method to obtain $\boldsymbol{W}$. We then solve an  
optimization problem termed as TV-TV minimization that takes $\boldsymbol{W}, \boldsymbol{Z}$ and $\boldsymbol{Y}$ as inputs to generate an improved version of $\boldsymbol{W}$ denoted by $\boldsymbol{\widehat{X}}$. This approach extends our original work \cite{vella2019tvtvminimization, Vella20-Robust}, where we considered single image super-resolution. In this paper, we generalize our original framework to fusion based HS images by considering the additional constraint $\boldsymbol{Y} = \boldsymbol{XR}$, which relates data from different modalities. Our extensive experiments show that the proposed method surpasses the current state-of-the-art for fusion HS super-resolution in terms of several reconstruction metrics. To the best of our knowledge, this is the first work that tackles the data consistency problem characterizing DL networks in the field of fusion based HS imaging super-resolution.

\section{Related Work}

In recent years, single-image super-resolution methods that process conventional RGB images have been extended to HS imaging. Traditional HS image super-resolution methods take a LRHS image as an input and output a HRHS image, without the aid of any other modality. Because of this, such methods have two shortcomings: the scaling factor is usually limited to four, and they increase the spectral resolution only. To overcome these shortcomings, several techniques fuse a LRHS image with a conventional HRMS image. Most methods are model-based and require explicit assumptions about the HRHS image, but more recent algorithms build on DL networks. The latter is still an emerging field with limited literature. We give a brief overview of both techniques since we utilize both in our work.

\mypar{Model-based methods} Classic model-based methods recover $\boldsymbol{X}$ by making explicit assumptions about it. For example, in \cite{Simoes2015tv}, it is assumed that $\boldsymbol{X}$ has a small TV norm. The work in \cite{Akhtar2014sparsespatiospectral} assumes that the spatial information of the HRHS image has a sparse representation that can be learned from a dictionary. 

A different line of work is based on matrix factorization methods \cite{Kawakami2011tfm, Yokoya2012cnmf}. These assume that the HS observations can be factorized into two matrices, one representing the spectral response of the material, and the other containing the proportion of materials at each pixel location. For example, \cite{Yokoya2012cnmf} proposed an algorithm that uses coupled non-negative matrix factorization (CNMF) and assumes no sparsity constraints. Even though matrix factorization approaches are popular, they ignore the spectral-spatial correlations. Motivated by this, more recent methods use tensor factorization. For example, \cite{dian2017nlstf} uses non-local sparse tensor factorization to decompose each cube of the HS image as a sparse tensor and dictionaries of three different modes. In addition, similar cubes are processed together to leverage the non-local self-similarities of the HS image. Such assumptions are common in other HS-related tasks, for example, tensor denoising \cite{gong2020}.

\mypar{Deep learning methods} DL has been widely applied to computer vision and image processing, e.g., \cite{MSLapSRN, Kim2016VDSR}, but its application to HS image super-resolution is still relatively recent \cite{Palsson2017fusion,Han2018ssfcnn, xie2019fusion, Sidrov2019Deephsi, Zhang2019hsir}. DL methods are computationally efficient during testing and offer superior performance. In reconstruction tasks, trained DL networks, however, cannot ensure that the output is consistent with the input \cite{Vella20-overcoming}. This is because the weights learned during training are applied to all test images, which makes it difficult for such networks to ensure that the observation model holds in the testing stage. The work in \cite{xie2019fusion} attempts to solve this problem by unfolding an iterative algorithm into a deep network that takes into consideration the low rankness and sparsity of the HRHS image. Also, \cite{qu2018unsupervised} proposed an unsupervised method consisting of two networks with an encoder-decoder structure used for representation learning. The first network takes the HRMS image as an input while the other network processes the LRHS image. To exploit spectral and spatial information, the learned weights are then shared before reconstructing the HRHS image. The proposed network also forces the representations of both inputs to follow a sparse Dirichlet distribution. Unsupervised networks solve a fundamentally harder task, thus their performance is naturally worse than supervised methods \cite{l2019handsonul}.
 

 \begin{table}[t]
   \centering
   \renewcommand{\arraystretch}{1}
   \caption{Average PSNR, SSIM, SAM, ERGAS and RMSE on all HS images from the Harvard dataset.}
   \begin{tikzpicture}
     \filldraw[color=black!10] (-2.4,-1.71) rectangle +(2.9,3.4);
     \filldraw[color=black!10] (0.73,-1.71) rectangle +(3.0,3.4);
     \draw node {
         \begin{tabular}{@{}llllll@{}}\toprule
           \textcolor{thisblue}{\textbf{Metric}} & 
           \textcolor{darkgreen}{\sf \textit{CNMF}} \cite{Yokoya2012cnmf}  & 		\textcolor{darkredc} {\sf \textit{\textbf{Ours}}}  & 			\textcolor{darkgreen}{\sf \textit{uSDN}} \cite{qu2018unsupervised}  &
           \textcolor{darkredc} {\sf \textit{\textbf{Ours}}} \\
           \midrule
           {{PSNR} } &41.381 & \textbf{41.574}  & 38.564 & \textbf{41.120}\\
           \midrule
           {SSIM} & \phantom{0}0.985 & \phantom{0}\textbf{0.986} & \phantom{0}{0.990}&  \phantom{0}\textbf{0.993}\\
           \midrule
           {SAM} & \phantom{0}3.951 &\phantom{0}\textbf{3.831} & \phantom{0}4.503&  \phantom{0}\textbf{3.441} \\
           \midrule
           {ERGAS} & \phantom{0}0.314 &\phantom{0}\textbf{0.311} & \phantom{0}0.563 & \phantom{0}\textbf{0.431}\\
           \midrule
           {RMSE} & \phantom{0}2.456&  \phantom{0}\textbf{2.407}& \phantom{0}2.753 & \phantom{0}\textbf{1.817}\\
           \bottomrule
         \end{tabular}
       };
   \end{tikzpicture}
   \label{tab:harvard}
   \vspace{-0.4cm}
 \end{table}

  \begin{table*}[t]
	\sffamily\selectfont
	\centering
	\renewcommand{\arraystretch}{1}
	\caption{Average PSNR, SSIM, SAM, ERGAS and RMSE on 12 test HS images used by MHF from the CAVE dataset.}
    \begin{tikzpicture}
    \filldraw[color=black!10] (-5.8,-1.71) rectangle +(3.05,3.4);
     \filldraw[color=black!10] (-2.5,-1.71) rectangle +(3.05,3.4);
     \filldraw[color=black!10] (0.79,-1.71) rectangle +(3.05,3.4);
     \filldraw[color=black!10] (4.1,-1.71) rectangle +(3.15,3.4);
     \draw node {
		\begin{tabular}{@{}lllllllll@{}}
			\toprule[1.1pt]
			\textcolor{thisblue}{\sf \textbf{Metric}} & 
			\textcolor{darkgreen}{\sf \textit{CNMF}} \cite{Yokoya2012cnmf} & 
			\textcolor{darkredc} {\sf \textit{\textbf{Ours}}} &
			\textcolor{darkgreen}{\sf \textit{NLSTF}} \cite{dian2017nlstf} &
			\textcolor{darkredc}{\sf \textit{\textbf{Ours}}}   &
			\textcolor{darkgreen}{\sf \text{uSDN}} \cite{qu2018unsupervised}  &
			\textcolor{darkredc} {\sf \textit{\textbf{Ours}}} &
			\textcolor{darkgreen}{\sf \textit{MHF-net}} \cite{xie2019fusion}   &
			\textcolor{darkredc} {\sf \textit{\textbf{Ours}}} 
			\\
			\midrule
			{PSNR} & 37.149 & \textbf{ 37.881} & 40.859 & \textbf{40.942} & 36.365 & \textbf{39.712 }& 37.507 & \textbf{38.151}\\
			\midrule
			{SSIM} &  $\phantom{0}$0.982 & $\phantom{0}$ \textbf{0.984 }&$\phantom{0}$ 0.991 &$\phantom{0}$\textbf{0.991}  &$\phantom{0}$0.979 &$\phantom{0}$\textbf{0.987} & $\phantom{0}$0.977 &$\phantom{0}$\textbf{0.982} \\
			\midrule
			{SAM} &$\phantom{0}$7.482 &$\phantom{0}$ \textbf{7.263}  &$\phantom{0}$ 4.723 &$\phantom{0}$\textbf{4.719}& $\phantom{0}$ 6.823&$\phantom{0}$\textbf{6.051} &$\phantom{0}$8.221  &$\phantom{0}$\textbf{7.946}\\
			\midrule
			{ERGAS} &$\phantom{0}$0.629 &$\phantom{0}$ \textbf{0.568}  & $\phantom{0}$ 0.384  &$\phantom{0}$\textbf{0.379}  & $\phantom{0}$ 0.716 & $\phantom{0}$\textbf{0.533}  &$\phantom{0}$0.590 & $\phantom{0}$\textbf{0.541}\\
			\midrule
			{RMSE} &$\phantom{0}$3.805 &$\phantom{0}$ \textbf{3.487}&$\phantom{0}$ 2.598  &$\phantom{0}$\textbf{2.568}   & $\phantom{0}$ 4.235 &$\phantom{0}$\textbf{2.877} & $\phantom{0}$3.845 &$\phantom{0}$\textbf{3.565}\\
			\bottomrule
		\end{tabular}
	};
	\end{tikzpicture}
	\label{tab:cave}
	\vspace{-0.4cm}
\end{table*}
 
\section{Proposed Framework}
\label{ourframework}

Our goal is to recover the HRHS image $\boldsymbol{X} \in \mathbb{R}^{M_0.N_0 \times S_0}$ from a LRHS image $\boldsymbol{Z} \in \mathbb{R}^{M.N \times S_0}$ and an RGB image $\boldsymbol{Y} \in \mathbb{R}^{M_0.N_0 \times S}$. To achieve this, we use the scheme depicted in Fig. \ref{fig:frameworkV2}. First, the inputs $\boldsymbol{Z}$ and $\boldsymbol{Y}$ are super-resolved into $\boldsymbol{W} \in \mathbb{R}^{M_0.N_0 \times S_0}$ via a fusion based HS image super-resolution method. We use a combination of both model and DL based methods and then post-process $\boldsymbol{W}$ via an additional block, TV-TV minimization, which enforces consistency for both the spatial and spectral measurements. 

\mypar{TV-TV minimization} The images $\boldsymbol{W}$, $\boldsymbol{Y}$ and $\boldsymbol{Z}$ are then simultaneously processed via TV-TV minimization, which generates an estimate $\boldsymbol{\widehat{X}}$ that satisfies the model in (\ref{eq:cond1}), is not too different from $\boldsymbol{W}$, and has a small total variation (TV) norm. More specifically, we solve

 \begin{equation}
\begin{array}{cl}{\underset{\boldsymbol{X} \in \mathbb{R}^{M_0.N_0 \times S_0}}{\operatorname{minimize}}} & {\|\boldsymbol{X}\|_{\mathrm{\text{TV}}}+\beta\|\boldsymbol{X} - \boldsymbol{W}\|_{\text{TV}}} \\  {\text {subject to}} & {\boldsymbol{AX}=\boldsymbol{Z}}\\ & {\boldsymbol{XR}=\boldsymbol{Y},}\end{array}
\label{eq:tv-tv}
\end{equation}
where $\boldsymbol{X} \in \mathbb{R}^{M_0.N_0 \times S_0}$ is the optimization variable and $\beta \geq 0$ is a trade-off parameter. We chose the TV norm since it is a widely used prior for image processing tasks. Our framework can be easily adapted to accommodate other priors. The TV-norm $\| \boldsymbol{X} \|_{\text{TV}}$ in (\ref{eq:tv-tv}) for 3D tensors is defined as the sum of the 2D TV-norms of each spectral band i.e., $\| \boldsymbol{X} \|_{\text{TV}} = \sum_{i=1}^{S_0}\left\|X_{s}\right\|_{\mathrm{TV}}$ where $X_s \in \mathbb{R}^{M_0 \times N_0}$ represents the band $s$ of $X$. The 2D TV norm of a vectorized image $x \in \mathbb{R}^{M_0. N_0}$ is defined as 
$$
\left\|x\right\|_{\mathrm{TV}}:=\sum_{i=1}^{M} \sum_{j=1}^{N}\big|D_{i j}^{v} x\big|+\big|D_{i j}^{h} x\big| = \|\boldsymbol{D}x\|_1,
$$
where $D^v_{ij}$ (resp. $D^h_{ij}$) is a row-vector that extracts the vertical (resp. horizontal) difference at pixel $(i, j)$ of $\boldsymbol{X}$, and $\boldsymbol{D}$ is the vertical concatenation of all the $D^v_{ij}$ and $D_{ij}^h$. 

To solve (\ref{eq:tv-tv}), we use the alternating direction method of multipliers (ADMM) \cite{boyd_distributed_2011}. First we introduce a set of auxiliary variables, $u_s$ and $v_s$, which we constrain as  $\boldsymbol{D}v_s = u_s$ and $x_s=v_s$ where $u_s$ (resp. $v_s$ and $x_s$) represents the vectorized $s$th spectral band of $u$ (resp. $v$ and $x$), as suggested in \cite{liu_depth_2015}, and dualize both constraints. Specifically, we rewrite (\ref{eq:tv-tv}) as
\begin{equation}
  \begin{array}[t]{ll}
    \underset{(u,\boldsymbol{X}),v}{\text{minimize}} &  \sum^{S_0}_{s=1} \big( \|u_s\|_1 + \beta \|u_s-\overline{w}_s\|_1 + \\ &
    \text{i}_{\{x_s:Ax_s=z_s\}}(x_s) \big)+ \text{i}_{\{\boldsymbol{X}:\boldsymbol{XR}=\boldsymbol{Y}\}}(\boldsymbol{X})\\ \\
        \text{subject to} & \boldsymbol{D}v_s=u_s, s = 1, \cdots , S_0 \\ 
        & v_s=x_s, s = 1, \cdots , S_0 \, ,
  \end{array}
  \label{eq:tvvec}
\end{equation}
where $\overline{w}_s := \boldsymbol{D}w_s \in \mathbb{R}^{M_0.N_0}$ and $\text{i}_F(u)$ is
an indicator function, i.e., $\text{i}_F(u) = 0$ if $u \in F$, and
$\text{i}_F(u) = +\infty$ otherwise. Once we obtain a solution ($u^\star, X^\star, v^\star$) of (\ref{eq:tvvec}), a solution of (\ref{eq:tv-tv}) is given by $X^\star = \begin{bmatrix} x^{\star}_1 & \cdots & x^{\star}_{S_0}\end{bmatrix}$.

\mypar{ADMM iterates} We use two dual variables $\lambda \in \mathbb{R}^{2 \cdot M \cdot N}$ and $\mu \in \mathbb{R}^{M \cdot N} $ and solve (\ref{eq:tvvec}) via the following ADMM iterates:
\begin{align}
  \boldsymbol{U}^{k+1} 
  &= 
  \underset{u_1 \cdots u_{S_0}}{\arg\min}\,\,\,
  \sum_{s=1}^{S_0} \|u_s\|_1 + \beta \|u_s-w_s\|_1 + 
  \notag
  \\  
  &\hspace{3.1cm}
  \big(\lambda_s^{k} - \rho  \boldsymbol{D}v_s^k\big)^T u_s 
  + \frac{\rho_s}{2} \|u_s\|^2_2
  \label{Eq:ADMMSubProbU}
  \\
  \boldsymbol{X}^{k+1} 
  &= 
  \underset{\boldsymbol{X}}{\arg\min}\,\,\,
  \mu^{k^T}\boldsymbol{X} + \frac{\rho}{2}
  \|\boldsymbol{X}-\boldsymbol{V}^k\|^2_2 
  +
  \text{i}_{\{\boldsymbol{X}:\boldsymbol{AX}=\boldsymbol{Z}\}}(\boldsymbol{X})
  \notag
  \\
  &\hspace{4.3cm}
  +
  \text{i}_{\{\boldsymbol{X}:\boldsymbol{XR}=\boldsymbol{Y}\}}(\boldsymbol{X})
  \label{Eq:ADMMSubProbX}
  \\
  \boldsymbol{V}^{k+1} 
  &= 
  \underset{v_1 \cdots v_{S_0}}{\arg\min}\,\,\,
  \sum_{s=1}^{S_0} -\lambda_s^{k^T} \boldsymbol{D}v_s - \mu^{k_s^T}v_s 
  + \frac{\rho}{2} \|x_s^{k+1}-v_s\|^2_2 
  \notag
  \\
  &\hspace{1.7cm}
  + \frac{\rho}{2} \|x_s^{k+1}-v_s\|^2_2 + \frac{\rho}{2} \|u_s^{k+1}- \boldsymbol{D}v_s\|^2_2 
  \label{Eq:ADMMSubProbV}
  \\
  \lambda^{k+1} 
  &= 
  \sum^{S_0}_{s=1} \lambda_s^k + \rho(u_s^{k+1}- \boldsymbol{D}v_s^{k+1}) 
  \\
  \mu^{k+1} 
  &= 
  \sum_{s=1}^{S_0} \mu_s^k + \rho(x_s^{k+1}- v^{k+1}_s)\,, 
  \label{Eq:lmult}
\end{align}
where $\rho > 0$ is the augmented Lagrangian parameter. Recall that the subscript $s$ represents the $s$th band of each of the hyperspectral image. Then the concatenation of each band gives the corresponding hyperspectral cube. Problem (\ref{Eq:ADMMSubProbU}) is solved by equating the element-wise derivative with respect to $u_s$ to zero, (\ref{Eq:ADMMSubProbX}) is the projection of a point onto a linear subspace, and (\ref{Eq:ADMMSubProbV}) can be solved by equating the first order derivative with respect to $v_s$ to zero. Thus, each subproblem has a closed form solution. 
Note that problems (\ref{Eq:ADMMSubProbU}) and (\ref{Eq:ADMMSubProbV}) can each be decomposed into $S_0$ independent problems that can be solved in parallel, which considerably speeds up the ADMM iterates.

\section{Experiments}
 \label{sec:experiments}

We now describe our testing procedure and present visual and quantitative results. Code to replicate our experiments is available online\footnote{https://github.com/marijavella/hs-sr-tvtv}. 

\mypar{Datasets} In this paper we run experiments on two popular HS imaging datasets, CAVE \cite{yasuma2010cave} and Harvard \cite{chakrabarti2011statistics}. The CAVE dataset consists of 32 HRHS images, each with dimensions $512 \times 512$ and 31 spectral bands. The spectral images are taken with a wavelength ranging between $400 \sim 700$nm at an interval of 10nm. The Harvard dataset consists of 50 HRHS images with dimensions $1392 \times 1040$ and 31 spectral bands each. Each spectral image is taken at 10nm intervals and the wavelength ranges between $420 \sim 720$nm. For this dataset, the top left 1024 $\times$ 1024 pixels were considered.

\mypar{Base methods} We use and compare against four state-of-the-art base methods: CNMF \cite{Yokoya2012cnmf}, NLSTF \cite{dian2017nlstf}, MHF \cite{xie2019fusion} and uSDN \cite{qu2018unsupervised}. The first and second methods are based on matrix and tensor factorization, respectively, while the last two use DL networks. The outputs from each base method were obtained using the default settings and parameters provided by the respective authors. We assume noiseless images and thus remove any added noise in any of the comparison methods.

\mypar{Experimental Settings} All images were first normalized between 0 and 1. For the CAVE dataset, we used the same 12 HS images/cubes tested with MHF while for the Harvard dataset, we use all the 50 available HS images/cubes. The HSLR images $\boldsymbol{Z}$ were obtained according to (\ref{eq:cond1}) with $\boldsymbol{A}$ representing the averaging operation over non-overlapping blocks of size $32 \times 32$, that is, as a downscaling operation with a factor of 32. The HRMS images $\boldsymbol{Y}$ were generated according to (\ref{eq:cond1}), with $\boldsymbol{R}$ representing either the camera spectral response of Nikon D700 \cite{qu2018unsupervised, dian2017nlstf} or an estimate as in \cite{Yokoya2012cnmf}. Details about the CSR function used in MHF \cite{xie2019fusion} are omitted in that paper but we used the matrix provided in the official repository\footnote{https://github.com/XieQi2015/MHF-net/tree/master/CMHF-net}. Since the CSR function varies in different methods, our aim is not to compare base methods against each other, but rather how our framework improves each of their outputs. 

For all base methods, we set $\beta = 1$ as the trade-off parameter in (\ref{eq:tv-tv}), and $\rho = 0.2$ as the augmented Lagragian parameter in (\ref{Eq:ADMMSubProbU})-(\ref{Eq:lmult}). The algorithm was terminated when the primal or dual residual was smaller than 0.001, or the number of iterations exceeded 120.
 
The performance of each method was assessed by the following quantitative measures: peak-signal-to-noise ratio (PSNR), structural similarity (SSIM) \cite{Wang2004ssim}, spectral angle mapper (SAM) \cite{Yuhas1992DiscriminationAS}, relative dimensionless global error in synthesis (ERGAS) \cite{Wald2002DataFD} and root mean squared error (RMSE).  

    \begin{figure}[t]
 	\centering
 	\begin{subfigure}[b]{0.25\textwidth}
 		\centering
 		\includegraphics[width=2.5cm, height=4cm]{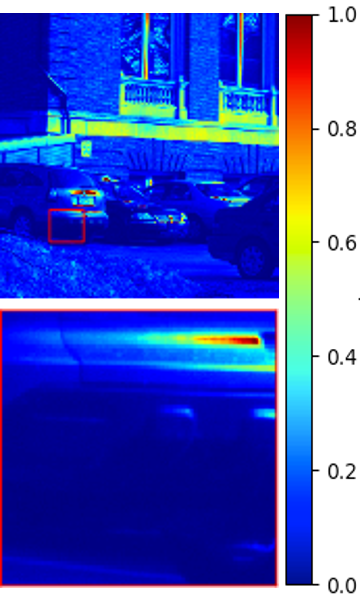}
 		\caption{GT}
 	\end{subfigure}
 	\begin{subfigure}[b]{2.5cm}
 		\centering
 		\includegraphics[width=2cm, height=4cm]{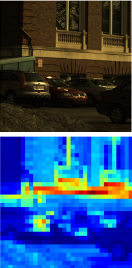}
 		\caption{RGB/LR}
 	\end{subfigure}
 	\begin{subfigure}[b]{0.11\textwidth}
 		\centering
 		\includegraphics[width=2cm, height=4cm]{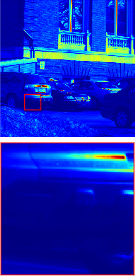}
 		\caption{CNMF}
 	\end{subfigure}
 	\begin{subfigure}[b]{0.11\textwidth}
 		\centering
 		\includegraphics[width=2cm, height=4cm]{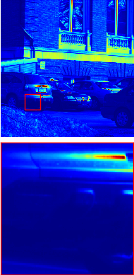}
 		\caption{Ours+CNMF}
 	\end{subfigure}
 	\begin{subfigure}[b]{0.11\textwidth}
 		\centering
 		\includegraphics[width=2cm, height=4cm]{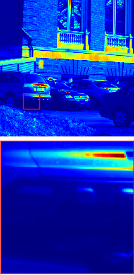}
 		\caption{uSDN}
 	\end{subfigure}
 	\begin{subfigure}[b]{0.11\textwidth}
 		\centering
 		\includegraphics[width=2cm, height=4cm]{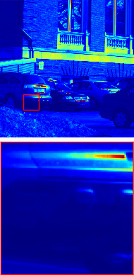}
 		\caption{Ours+uSDN}
 	\end{subfigure}
 	\caption{Reconstruction of band 15 from a sample image from the Harvard dataset using CNMF and uSDN as base methods.}
 	\label{fig:harvard_results}
 	\vspace{-0.5cm}
 \end{figure}

\mypar{Experiments on the CAVE and Harvard datasets} Table \ref{tab:harvard} (resp. \ref{tab:cave}) shows the average PSNR, SSIM, SAM, ERGAS and RMSE of \cite{Yokoya2012cnmf, dian2017nlstf, xie2019fusion, qu2018unsupervised} and of our algorithm using those base methods as input, on the Harvard and CAVE datasets. Larger PSNR and SSIM and smaller ERGAS, SAM and RMSE indicate images of better quality. The results clearly show that our framework consistently achieves better metric scores on all the datasets and methods considered, confirming that we are able to obtain better spatial and spectral resolution by enforcing measurement consistency. For example, for the CAVE dataset, the SAM gains range between 0.004 and 0.772, while the improvements of the ERGAS range between 0.005 and 0.183. Also, note that our method provides larger gains for DL based methods, which cannot easily enforce consistency during testing.

\begin{figure}[t]
 	\centering
 	\begin{subfigure}[b]{0.25\textwidth}
 		\centering
 		\includegraphics[width=2.4cm, height=4.3cm]{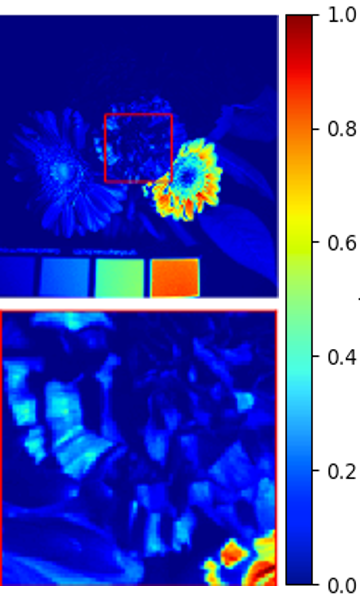}
 		\caption{GT}
 	\end{subfigure}
 	\begin{subfigure}[b]{2.5cm}
 		\centering
 		\includegraphics[width=2cm, height=4.2cm]{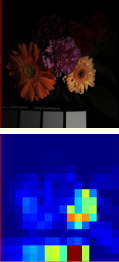}
 		\caption{RGB/LR}
 	\end{subfigure}
 	\begin{subfigure}[b]{0.11\textwidth}
 		\centering
 		\includegraphics[width=2cm, height=4cm]{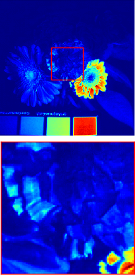}
 		\caption{MHF}
 	\end{subfigure}
 	\begin{subfigure}[b]{0.11\textwidth}
 		\centering
 		\includegraphics[width=2cm, height=4cm]{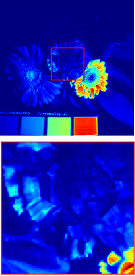}
 		\caption{Ours+MHF}
 	\end{subfigure}
 	\begin{subfigure}[b]{0.11\textwidth}
 		\centering
 		\includegraphics[width=2cm, height=4cm]{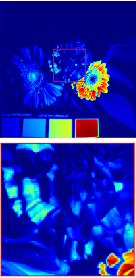}
 		\caption{uSDN}
 	\end{subfigure}
 	\begin{subfigure}[b]{0.11\textwidth}
 		\centering
 		\includegraphics[width=2cm, height=4cm]{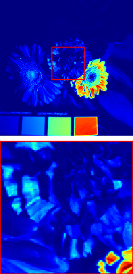}
 		\caption{Ours+uSDN}
 	\end{subfigure}
 	\begin{subfigure}[b]{0.11\textwidth}
 		\centering
 		\includegraphics[width=2cm, height=4cm]{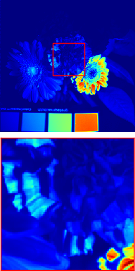}
 		\caption{CNMF}
 	\end{subfigure}
 	\begin{subfigure}[b]{0.11\textwidth}
 		\centering
 		\includegraphics[width=2cm, height=4cm]{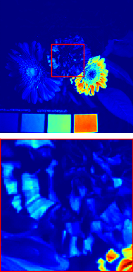}
 		\caption{Ours+CNMF}
 	\end{subfigure}
 	\begin{subfigure}[b]{0.11\textwidth}
 		\centering
 		\includegraphics[width=2cm, height=4cm]{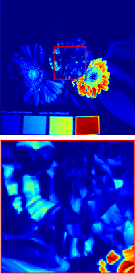}
 		\caption{NLSTF}
 	\end{subfigure}
 	\begin{subfigure}[b]{0.11\textwidth}
 		\centering
 		\includegraphics[width=2cm, height=4cm]{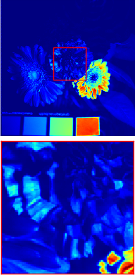}
 		\caption{Ours+NLSTF}
 	\end{subfigure}
 	\caption{Reconstruction of band 15 of the \textit{flowers} image from the CAVE dataset for all the base methods considered.}
 	\label{fig:caveresults}
 	\vspace{-0.5cm}
 \end{figure}

 We also present visual results, using the jet colormap for easier visualizations, in Figs. \ref{fig:harvard_results} and \ref{fig:caveresults} on a sample image from each dataset. The first subfigure a) of each figure contains the ground truth (GT) image and a patch extracted from it, while the second subfigure b) presents the RGB and LR image, respectively. Note that the RGB image corresponds to the one used in MHF. Other networks have a different RGB image due to a different CSR function. The rest of the subfigures contain the output from a base method followed by the output from our method. This is repeated for each base method. 
 
 The superiority of our method can be observed from the extracted patches. The base methods tend to blur out details in the image, while our method preserves the original measurements in the outputs. Although the proposed algorithm improves the quality of the outputs of both model-based and learning-based methods in several metrics, one shortcoming is the additional computation it requires. For example, to upsample a $16 \times 16 \times 31$ HS image with a $32 \times$ factor and a $512 \times 512 \times 3$ RGB image, MHF (supervised method) requires 25 seconds, uSDN (unsupervised method) requires 660 seconds while our method requires 840 seconds.

\section{Conclusions}

We proposed a post-processing step for the fusion of RGB and HS images that can work with either classic or DL based methods. Results show that by guaranteeing consistency in the spatial and spectral measurements, the image quality is improved. Moreover, our method can easily work with different downscaling operators and camera spectral response functions, unlike DL networks that are trained for specific operators. The basis of our framework is TV-TV minimization, which we solve with an ADMM-based algorithm. Future lines of work include using different regularizers and unrolling our algorithm in a neural network to improve reconstruction time.

\section{Acknowledgements}
Work supported by Royal Society and NSFC (IEC/NSFC/181255), EPSRC 
(EP/T026111/1), National Key R\&D Program of China (Grant 
No.2018YFE0207600), the Natural Science Foundation of China 
(61911530216) and the Beijing Natural Science Foundation (L202019).

\bibliographystyle{IEEEbib}
\bibliography{refs}

\end{document}